\documentclass[aps,prx,amsmath,amssymb,floatfix,twocolumn]{revtex4-1}
\usepackage{parskip} 
\usepackage{tikz} 
\usepackage{graphicx,ulem}
\usepackage{subfigure}
\usepackage{amsmath}
\usepackage{bbold}
\usepackage{slashed}
\usepackage{amssymb}
\usepackage{indentfirst}
\usepackage{braket}
\usepackage{array}
\usepackage[colorlinks]{hyperref}
\usepackage{float}
\usepackage[margin=1in]{geometry}

\usepackage{natbib}

\begin{document}
\preprint{APS/123-QED}

\title{Decoupling the electronic gap from the spin Chern number in disordered higher-order topological insulators}
\author{Alexander C. Tyner$^{1,2}$}
\author{Cormac Grindall$^{3}$}
\author{J. H. Pixley$^{3,4}$}

\affiliation{$^{1}$Nordita, KTH Royal Institute of Technology and Stockholm University 106 91 Stockholm, Sweden}
\affiliation{$^2$Department of Physics, University of Connecticut, Storrs, Connecticut 06269, USA}
\affiliation{$^{3}$Department of Physics and Astronomy, Center for Materials Theory, Rutgers University, Piscataway, New Jersey 08854, USA}
\affiliation{$^{4}$Center for Computational Quantum Physics, Flatiron Institute, 162 5th Avenue, New York, NY 10010}

\date{\today}

\begin{abstract} 
In two-dimensional topological insulators, a disorder induced topological phase transition is typically identified with an Anderson localization transition at the Fermi energy. However, in higher-order, spin-resolved topological insulators it is the spectral gap of the spin-spectrum, in addition to the bulk mobility gap, which protects the non-trivial topology of the ground state. In this work, we show that these two gaps, the bulk electronic and spin gap, evolve distinctly upon introduction of disorder. This decoupling leads to a unique situation in which an Anderson localization transition occurs below the Fermi energy at the topological transition. Furthermore, in the clean limit the bulk-boundary correspondence of such higher-order insulators is dictated by crystalline protected topology, coexisting with the spin-resolved topology. By removing the crystalline symmetry, disorder allows for isolated study of the bulk-boundary correspondence of spin-resolved topology for which we demonstrate the absence of protected edge and corner modes in the Hamiltonian
and yet the edge modes in the eigenstates of the projected spin operator survive. Our work shows that a non-zero spin-Chern number, in the absence of a non-trivial $\mathbb{Z}_{2}$ index, does not dictate the existence of protected edge modes, resolving  a fundamental question posed in 2009.  
\end{abstract}

\maketitle
\par 
\section{Introduction}
Understanding the robustness of topological properties to weak 
perturbations in electronically gapped systems 
has remained a central question at the forefront of research in condensed matter physics\cite{Klitzing,TKNN,Kane2005,bernevig2006quantum,Hassan2010,Qi2011}.
The destruction of topological properties upon closing the electronic gap is a fundamental concept underlying this phenomena in a wide range of physical systems such as a two-dimensional electron gas in a magnetic field (e.g. the integer\cite{Klitzing} and fractional\cite{laughlin,eisenstein1990fractional} quantum Hall effects) and topological band structures such as Chern\cite{haldane} and topological insulators\cite{Kane2005}.
Upon the introduction of weak disorder (that does not break any protecting symmetry such as time reversal in the case of $\mathbb{Z}_2$ topological insulators), this electronic  gap closes (being filled in by Anderson localized Lifshitz states\cite{thouless1970anderson,klopp2002weak}) but so long as the mobility gap remains open, the topological properties can be rigorously proven to remain intact\cite{Kane2005,FuKane,bernevig2006quantum,ZhangFT,Xu2006,Ryu2007,Ryu2009,Shindou2009,ryu2010topological,Hassan2010,Schubert2012,Leung2012,Kobayashi2013,Morimoto2014,katsura2016,Chiu2016}.
As a consequence, the bulk mobility gap is %
always
used as a metric of how ``protected" the ground state topology is to external perturbations\cite{Xu2006,Hassan2010,Qi2011}.

In quantum spin Hall insulators this same paradigm has been widely adopted. However,  the presence of non-trivial spin-resolved topology requires the preservation of a gap in both the electronic and spin spectrum~\cite{Prodan2009}. 
A gapped spin spectrum is essential as the spin-Chern number is a relative invariant between the up and down spin sectors~\cite{bernevig2006quantum,Prodan2009,FuKane};
if the spin-spectrum is gapless the up and down sectors can hybridize and trivialize the invariant. 
In the presence of disorder, these gaps are converted into mobility gaps, i.e. their spectrum is soft but their eigenstates are localized such that they remain insulators with regards to transport.
\par
In many prominent cases the bulk spin mobility gap mirrors that of the bulk electronic mobility gap in the presence of disorder. This is true of $\mathbb{Z}_{2}$ non-trivial spin-Hall insulators\cite{bernevig2006quantum,Ryu2007,Prodan2009}. It is also true in higher-order or crystalline topological insulators, provided the disorder does not violate the crystalline symmetries protecting the bulk topology~\cite{Li2020,Hu2021,Wang2020,Wang2021,Wangprb2021,Wang2020,shen2023disorder}. It has therefore been common practice to ignore the spin-gap altogether and only study  the electronic spectrum. 
\par
\begin{figure}
    \centering
    \includegraphics[width=8cm]{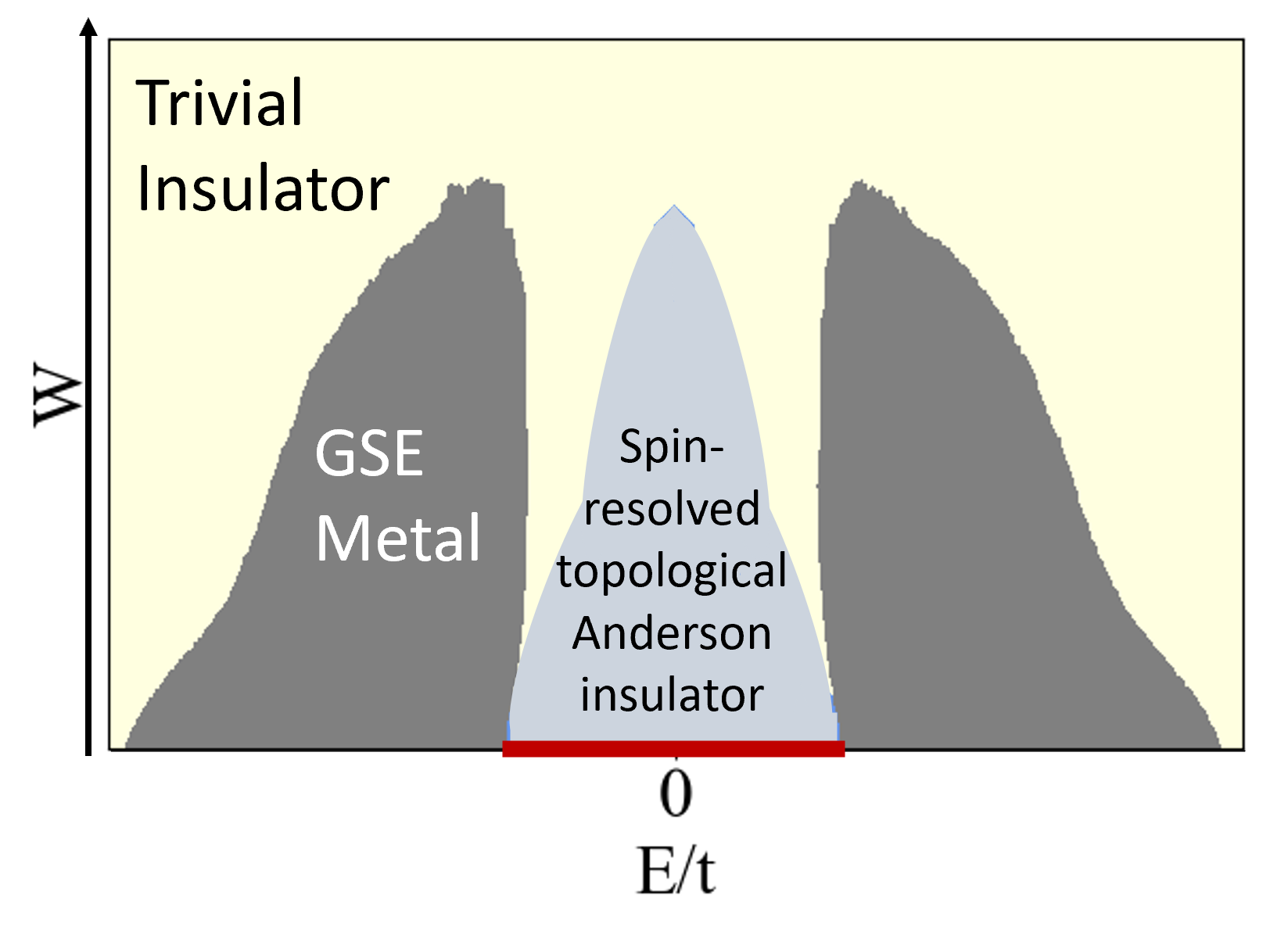}
    \caption{Schematic bulk phase diagram of the disordered HOTI Hamiltonian in Eq.~\eqref{eqn:fullH} as a function of energy $E$ and strength of the random on-site disorder $W$. The higher-order topological phase, which is only stable in the clean limit ($W=0$) is marked by a bold red line. The metallic bands at finite energy (shown in dark grey) are described by the Gaussian symplectic ensemble (GSE)  of random matrix theory\cite{levelstats}, which localizes at sufficiently large disorder without closing the mobility gap between the two bands. When the Fermi energy is placed within the spin gap (in the light grey region) the spin Chern number is quantized and the system exhibits a spin-resolved Anderson topological insulator (TI). 
    The line separating the light yellow and light blue regions is the spin mobility edge that is decoupled from the electronic mobility edge that separates the grey and yellow regions. When all of the bulk states localize, the spin mobility gap closes and the spin-Chern number is no longer quantized.
    }
    \label{fig:c4phase}
\end{figure}

 
Here we show that in higher-order topological insulators when the disorder \emph{does not} preserve the symmetry protecting the topology it is essential to study both the electronic and spin spectrum concomitantly. By generalizing the typical density of states to the spin spectrum %
we show that it is possible to decouple the fate of the electronic and spin mobility gaps as a function of disorder. 
While the disorder driven topological spin Hall-to-trivial insulator phase transitions can be identified by the spin gap closing at the Fermi energy, in direct contrast with all of the other paradigmatic examples we have previously discussed, the electronic mobility gap does not close at the Fermi energy. Instead, we provide strong numerical evidence through simulations on large system sizes that an Anderson localization transition takes place below the Fermi energy, leading to the  phase diagram seen in Fig.~\eqref{fig:c4phase}, which is fundamentally distinct from its lower order topological counterparts.
\par
\begin{figure*}
    \centering
    \includegraphics[width=16cm]{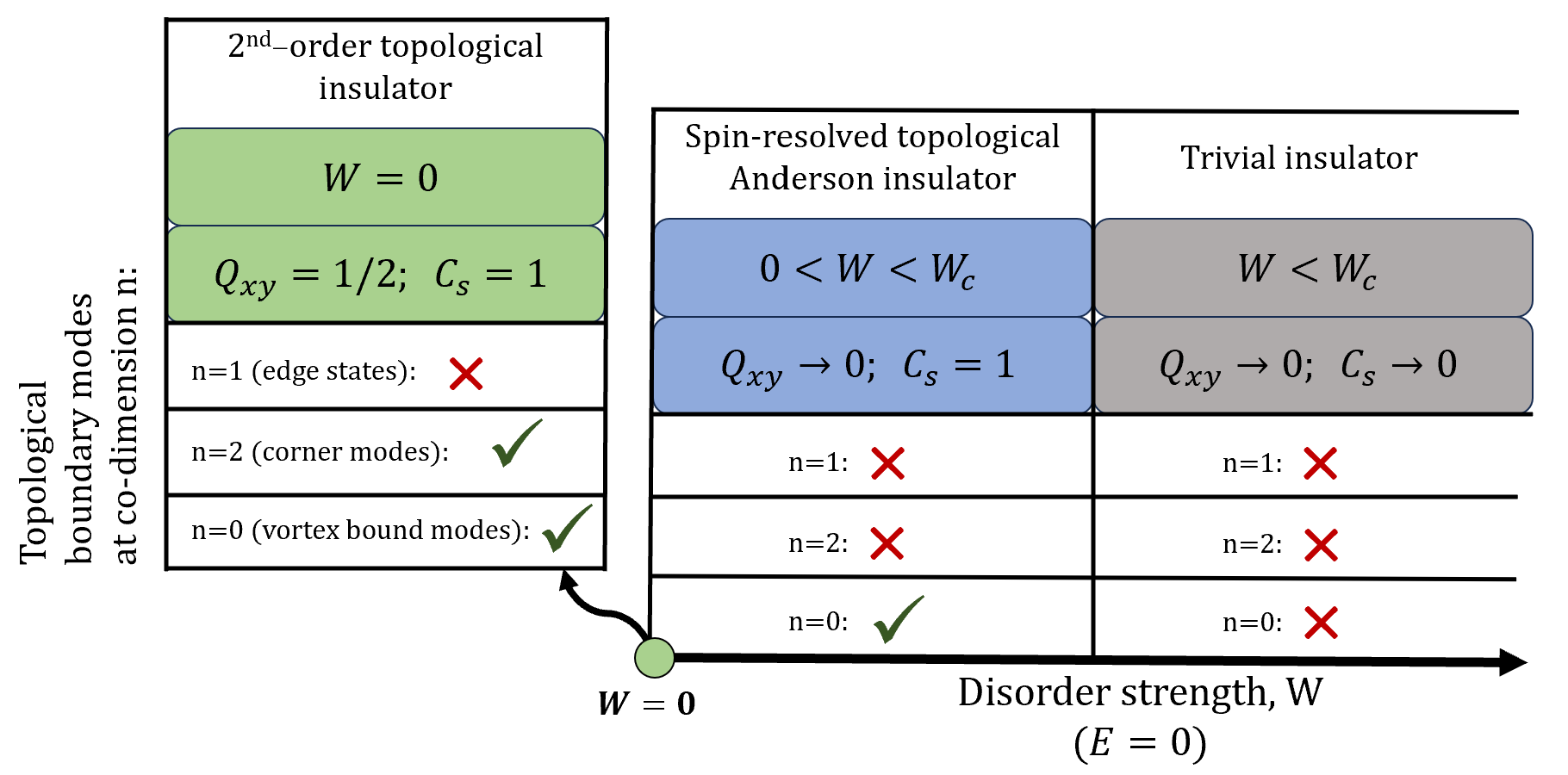}
    \caption{Evolution of bulk and surface topological properties of a spinful second order topological crystalline insulator at $E=0$ as a function of of random on-site disorder strength.
    }
    \label{fig:intro}
\end{figure*}

 
A major consequence of  decoupling  the electronic and spin mobility gaps must appear in the nature of edge modes as a result of the
bulk-boundary correspondence. 
In the clean limit, the bulk supports both spin-resolved topology and higher-order crystalline topology. Importantly, the bulk-boundary correspondence has been attributed almost entirely to the higher-order crystalline topology. However, by removing the protecting crystalline symmetries through the inclusion of disorder, as detailed schematically in Fig. \eqref{fig:intro}, we are able to localize the electronic surface states. As a result, we can  isolate and study the bulk-boundary correspondence arising due to the non-trivial spin-resolved topology. Interestingly, topologically protected edge/corner modes are shown to be absent in the states of the disordered Hamiltonian this case. Nonetheless, the PSO does yield well defined one-dimensional dispersing edge modes localized to their boundary provided the spin Chern number remains quantized.
In addition, the electronic spectrum is shown to support topological surface modes only on an inserted zero dimensional defect~\cite{QiSpinCharge,SpinChargeVishwanath}. 
\par
Importantly, in the landmark work on the spin Chern number\cite{Prodan2009}, Prodan noted ``the existence of spin Chern numbers does not automatically imply the existence of chiral edge modes." Our result serves as a proof that the existence of a spin-Chern number, in the absence of a non-trivial $\mathbb{Z}_{2}$ index, does \emph{not} dictate the existence of protected states. 
%


The remainder of the paper is organized as follows. In Sec. \eqref{Sec:ModelApproach} the tight-binding model that we study is introduced as well as details of how disorder will be modeled. In Sec. \eqref{sec:Clean} the electronic structure and topological properties of the tight-binding model in the clean limit are given. In Sec. \eqref{sec:observables} the computational tools utilized to examine the electronic and and topological properties of the tight-binding model in the presence of disorder are defined. In Sec. \eqref{sec:disorder} the effects of disorder are investigated, first with respect to the bulk and surface electronic spectrum in Sec. \eqref{sec:ElectronicDisorder}. Subsequently in Sec. \eqref{sec:SpinDisorder}, the effects of disorder on the bulk and surface spin-spectrum as well as topology of the spin-spectrum are explored. Finally before concluding in Sec. \eqref{sec:Summary}, we examine the use of magnetic flux tubes to probe the bulk topology of the spin-spectrum via the electronic spectrum in Sec. \eqref{sec:flux}.

\begin{figure}
\includegraphics[width=8cm]{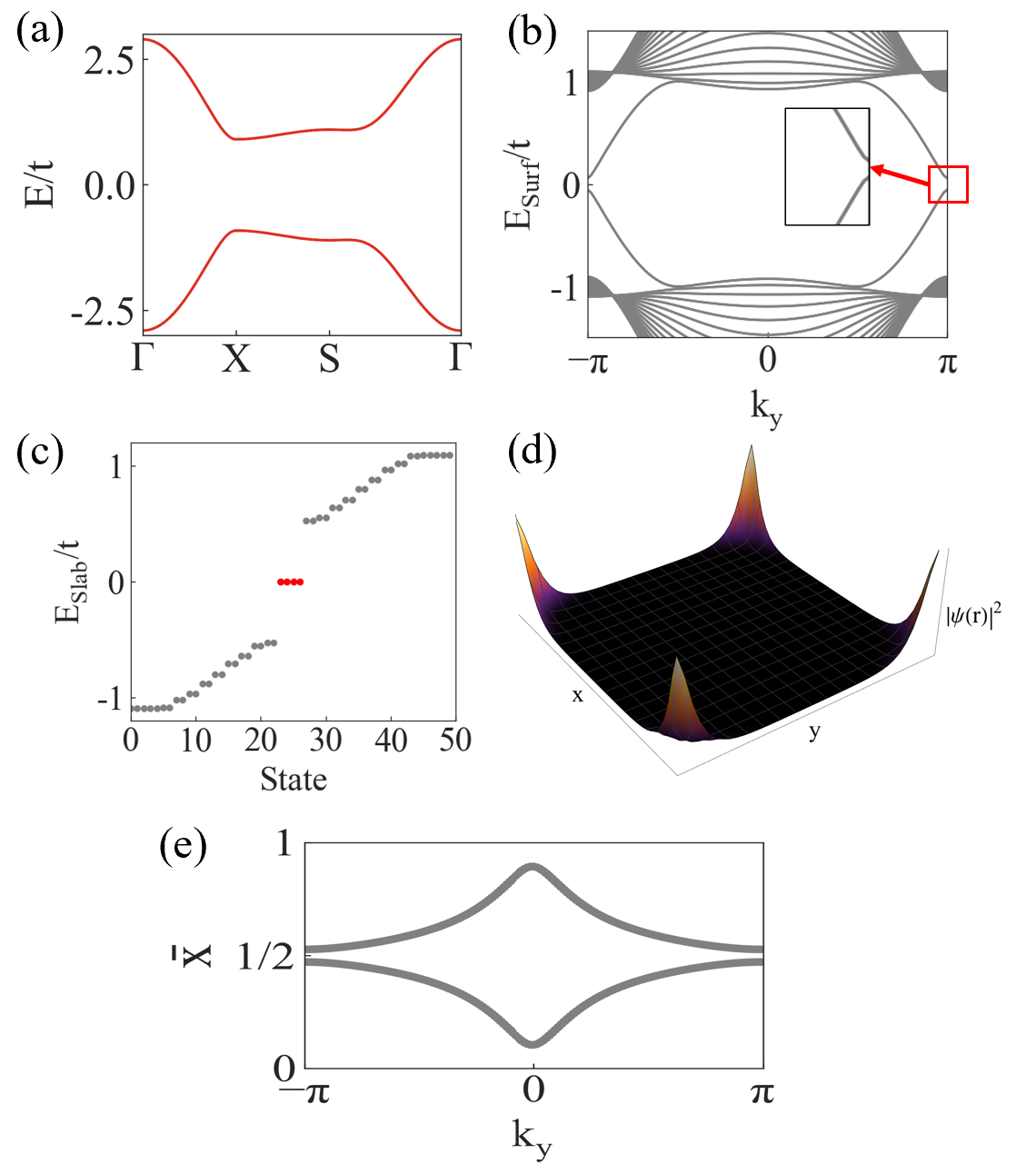}
\caption{
{\bf Properties of the HOTI band structure}:
(a) Band structure along high-symmetry path for tight-binding model given by Eq.~\eqref{eq:Model}. (b) Spectra considering a slab of 40 unit cells along the $\hat{x}$ direction and periodic boundary conditions along $\hat{y}$. Inset shows enlarged view of spectral gap at $k_{y}\pm \pi$. (c) Spectra of 50 lowest lying states when applying open-boundary conditions along both principal axes in a systems of $50 \times 50$ unit cells. Localization of degenerate mid-gap states shown in red is detailed in (d), demonstrating that they are corner localized. (e) Wannier center charge spectra for occupied bands. The spectra is gapped, disallowing assignment of non-trivial topology.
%
} 
\label{fig:CleanC4Model}
\end{figure}

\section{Model, symmetries, and approach}\label{Sec:ModelApproach}
We consider a quintessential tight-binding model, which is a spinful generalization of the famous Benalcazar-Bernevig-Hughes model~\cite{Benalcazar61,BenalacazarCn} to describe the HOTI band structure. We introduce a random onsite potential $V({\bf r})$ to model quenched disorder realizing the Hamiltonian, 
 \begin{equation}
     H=\sum_{{\bf k}}\psi_{{\bf k}}^{\dag}H_0({\bf k})\psi_{{\bf k}}+\sum_{{\bf r}}\psi_{{\bf r}}^{\dag}V({\bf r})\psi_{{\bf r}},
     \label{eqn:fullH}
 \end{equation}
 where $\psi_{\bf r}$ is a four component spinor at site ${\bf r}$ (or Bloch momentum ${\bf k}$). The Hamiltonian belongs to class AII\cite{Ryu2009,ryu2010topological,Chiu2016} (symplectic class), which hosts a stable delocalized phase at weak disorder in two-dimensions. 
 
 The Bloch Hamiltonian $H_0({\bf k})$ takes the form: 
\begin{multline}\label{eq:Model}
    H_{0}(\mathbf{k})/t=\sin k_{x} \sigma_{1}\otimes \tau_{1} + \sin k_{y}\sigma_{1}\otimes \tau_{2} \\+ M(\mathbf{k})\sigma_{3}\otimes \tau_{0} + \Delta(\mathbf{k})\sigma_{2}\otimes \tau_{0},
\end{multline}
 where t has units of energy, $\sigma_{0,1,2,3}(\tau_{0,1,2,3})$ are the $2\times 2$ identity matrix and three Pauli matrices respectively, operating on the spin (orbital) indices. We further define $M(\mathbf{k})=(\cos k_{x}+ \cos k_{y} + \gamma)$ and $\Delta(\mathbf{k})=t_{1}(\cos k_{x}- \cos k_{y})$, fixing $\gamma=0.9, \; t_{1}=0.05t$. Time-reversal symmetry $\mathcal{T}$, is generated by $\mathcal{T}^{\dagger}H^{*}(-\mathbf{k})\mathcal{T}=H(\mathbf{k})$, where $\mathcal{T}=i\tau_{3}\otimes \sigma_{1}$, such that $\mathcal{T}^{2}=-1$. 

 At each site, we model the quenched disorder as a random onsite  potential $V({\bf r})$. We sample the potential $V({\bf r})$ from a Gaussian distribution with zero mean and variance $W^2$, and hence $W$ characterizes the strength of disorder. Throughout this work, $W$ will be expressed in units of $t$.

In the following we present a detailed numerical study of the model defined in Eq.~\eqref{eqn:fullH} through a combination of exact diagonalization that reaches up to linear system sizes $L=50$ and the kernel polynomial method (KPM)~\cite{kpm} to study the average and typical density of states up to linear system sizes of $L=1000$. Before describing the quantities and metrics we use to determine each phase of the system we begin with a review of the model in the clean limit (i.e. $V({\bf r})=0$).

\begin{figure}
\includegraphics[width=8cm]{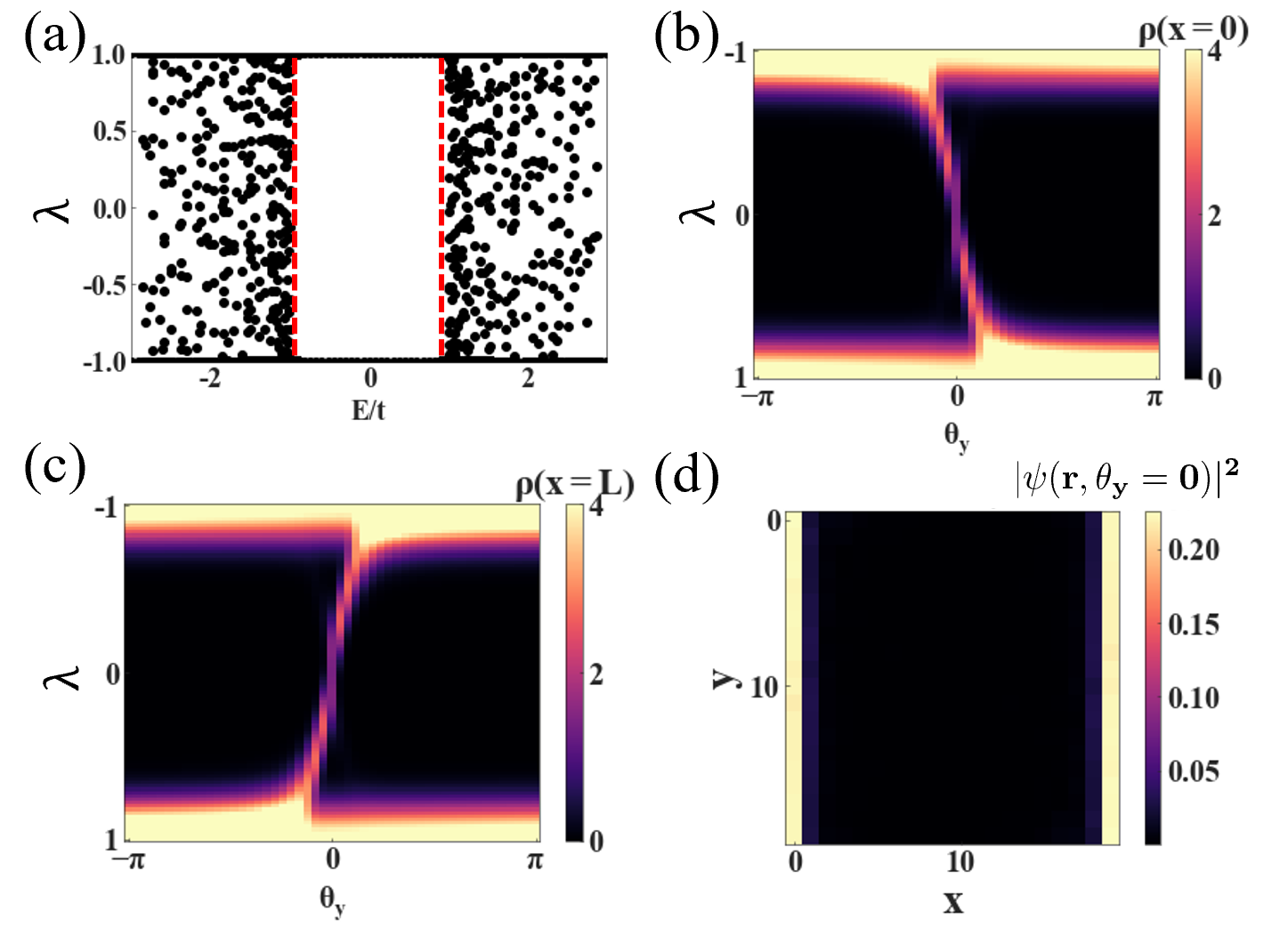}
\caption{
{\bf Properties of the PSO band structure}:
(a) Spectra of the PSO as a function of the Fermi energy on an $L=20$ system size with periodic boundary conditions. Red dashed line marks region of energy supporting a gapped PSO spectra. (b)-(c) The PSO on an $L=20$ size system with open boundary conditions along $x$ and twisted boundary conditions along $y$, for twist $\theta_{y}$, is considered. The average local density of states on the (b) $x=0$ and (c) $x=L$ surfaces as a function of the twist angle demonstrating chiral edge states. (d) The real-space distribution of the eigenstates nearest to $\lambda=0$ for twist angle $\theta_{y}=0$, demonstrating localization along the edges.
} 
\label{fig:CleanC4ModelPSO}
\end{figure}

\subsection{Clean limit}\label{sec:Clean}
 \par 
In the clean limit, Eq.~\eqref{eq:Model} supports four-fold rotational symmetry generated by, $C_{4}^{\dagger}H_{0}(k_{x},k_{y})C_{4}=H_{0}(k_{y},-k_{x})$, where $C_{4}=e^{-i\frac{\pi}{4}\tau_{0}\otimes \sigma_{3}}e^{-i\frac{\pi}{2}\tau_{3}\otimes \sigma_{3}}$. Additionally, a chiral symmetry, $S^{\dagger}H_{0}(\mathbf{k})S=-H_{0}(\mathbf{k})$ exists generated by $S=\tau_{1}\otimes \sigma_{3}$. In the presence of these additional symmetries the model falls under the category of HOTI.
\par 
The bulk band structure along the high-symmetry path is shown in Fig. \eqref{fig:CleanC4Model}(a) detailing that the model is an insulator for our choice of parameters. Considering a cylindrical geometry, such that $k_{y}$ remains a good quantum number, we find the spectra in Fig. \eqref{fig:CleanC4Model}(b) detailing the absence of gapless states on the $x$ edge. Finally, placing the system in a slab geometry of linear system size $L=50$ with open boundary conditions along $x$ and $y$ we find the spectra shown in Fig. \eqref{fig:CleanC4Model}(c), displaying four zero energy states. The localization of these states at the corners of the sample is shown in Fig. \eqref{fig:CleanC4Model}(d).
\par
The bulk-boundary correspondence for the corner bound states has been established in prior works through computation of the nested Wilson loop\cite{Benalcazar61,Schindlereaat0346}. The Wilson loop along direction $k_{x}$ as a function of $k_{y}$, is defined as,
\begin{equation}\label{eq:WCCs}
    W_{x}(k_{y})=\mathcal{P}\text{exp}\left[i\oint A_{occ,x}(\mathbf{k})dk_{x}\right],
\end{equation}
where $A_{occ,x}(\mathbf{k})= -i\bra{\Psi_{occ}}\partial_{k_{x}}\ket{\Psi_{occ}}$ is the Berry connection for the occupied subspace. Upon integration, the Wilson loop can be written in the form, 
\begin{equation}
    W_{x}(k_{y})=e^{iH_{W,x}(k_{y})}.
\end{equation}
The Wannier center charges (WCCs), $\bar{x}(k_{y})$, follow as eigenvalues of $H_{W,x}(k_{y})/\pi$ and are used to determine both the Fu-Kane $\mathbb{Z}_{2}$ index and  bulk-boundary correspondence for states on the $x$ edge. The WCC spectra for Eq.~\eqref{eq:Model} is shown to be gapped in Fig. \eqref{fig:CleanC4Model}(e), indicating a \emph{trivial $\mathbb{Z}_{2}$ index} and lack of gapless topological edge states. The nested Wilson loop follows as the Wilson loop of the Wannier Hamiltonian, $H_{W,x}(k_{y})$, and dictates that the corner modes are topologically protected by the presence of quantized quadrupole moment $Q_{xy}=1/2$\cite{Benalcazar61}, placing the system in the category of HOTIs with ``bulk-corner" correspondence\cite{Ezawa2020}. The quadrupole moment is in turn protected by the presence of chiral symmetry.  The symmetry protection of the quadrupole moment has been well documented in Refs. \cite{Benalcazar61,Li2020,Hu2021}. 

\par 
Coexisting with the symmetry protected topology of Eq.~\eqref{eq:Model}, is non-trivial spin-resolved topology. The spin-resolved topology is identified by a quantized spin-Chern number, $\mathcal{C}_{s}=(\mathcal{C}_{\uparrow}-\mathcal{C}_{\downarrow})/2$, where $\mathcal{C}_{\uparrow,\downarrow}$ is the Chern number defined for the spin-up and spin-down eigenstate which constitute a Kramers pair. To compute the spin-Chern number we must isolate the spin up(spin down) subsectors of the occupied Kramers pairs. This is accomplished through construction of the projected spin operator (PSO)\cite{Prodan2009}. The projected spin operator takes the form
$
\hat{\mathcal{S}}=P(\mathbf{k})\hat{s}P(\mathbf{k}),
$
where $P(\mathbf{k})$ is the projector onto occupied bands and $\hat{s}$ is a preferred spin axis. In the presence of spin-rotation symmetry it is simple to verify that the eigenvalues of the PSO are fixed as $\pm 1$. The positive(negative) branches of the spectra corresponding to the spin-up(spin-down) eigenstates for which the Chern number, $\mathcal{C}_{\uparrow}(\mathcal{C}_{\downarrow})$, can be computed. 
\par 
If additional terms violating the spin-rotation symmetry are introduced, the eigenvalues of the PSO adiabatically deviate from $\pm 1$. So long as a gap in the spectra of the PSO remains present, referred to as the spin-gap (labeled $\Delta_s$), $\mathcal{C}_{\uparrow}(\mathcal{C}_{\downarrow})$ can be unambiguously computed. Thus, the spin gap is as fundamental to the problem as the energy gaps. In the presence of disorder, these are converted into energy and spin mobility gaps that have to be determined from quantities that are not self-averaging.
\par
For Eq.~\eqref{eq:Model}, a gap in the PSO is present fixing the Fermi energy $E/t$=0 for the choice, $\hat{s}=s_{z}=\sigma_{3}\otimes \tau_{3}$, allowing for calculation of the spin-Chern number. The eigenvalues of the spin-spectra, as a function of the Fermi energy are shown in Fig. \eqref{fig:CleanC4ModelPSO}(a), demonstrating that the PSO is gapped for all values of the Fermi energy which fall within the electronic bulk-gap. 
\par 
Computation of the Chern number for the occupied eigenstates of the PSO ($\mathcal{C}_{\downarrow})$ is accomplished via modification of the coupling matrix method introduced in Ref. \cite{zhang2013coupling}, replacing the occupied eigenfunctions of the Hamiltonian with the eigenfunctions corresponding to the negative eigenvalues of the PSO. The results establish $\mathcal{C}_{s}=1$ for Eq.~\eqref{eq:Model}, while we emphasize that the $\mathbb{Z}_2$ index is trivial. 
\par
In correspondence with the presence of a quantized Chern number, the spectral density of the PSO  on the surface can be computed for open boundary conditions along $x$ and twisted boundary conditions along $y$, $\psi(x,y+L)=e^{i\theta_{y}}\psi(x,y)$. The results in Fig. \eqref{fig:CleanC4ModelPSO}(b) and Fig. \eqref{fig:CleanC4ModelPSO}(c) display the expected spectral flow. Fixing the twist angle, $\theta_{y}=0$, we inspect the real space distribution of the eigenstates at $\lambda=0$ in Fig. \eqref{fig:CleanC4ModelPSO}(d) that demonstrates that these gapless modes are bound to the open edges. 
\par 
The analysis of the Bloch Hamiltonian defined in Eq. \eqref{eq:Model} and presented in this section has established the coexistence of (I) symmetry-protected higher-order topology and (II) spin-resolved topology when the Fermi energy is placed within the bulk-gap. The presence of an electronic bulk gap is visible in Fig. \eqref{fig:CleanC4Model}(a). The electronic gap is marked by red-dashed lines in Fig. \eqref{fig:CleanC4ModelPSO}(a) demonstrating that within this region the PSO is correspondingly gapped. The bulk-boundary correspondence of the symmetry protected higher-order topology has been established through identification of a gap in the surface spectra in Fig. \eqref{fig:CleanC4Model}(b) and subsequent identification of zero-energy corner localized states as shown in Fig. \eqref{fig:CleanC4Model}(c)-(d). The bulk-boundary correspondence of the PSO as a Chern insulator has been established via analysis of the spectral flow upon imposition of twisted boundary conditions as seen in Fig, \eqref{fig:CleanC4ModelPSO}(b)-(c). In the following sections we will address whether the gap in the bulk electronic and spin-spectrum can be decoupled through introduction of disorder and the corresponding impact on the fate of topological boundary modes.

\subsection{Probes of the disordered model}
\label{sec:observables}
We now define all quantities computed to construct a complete phase diagram in the presence of disorder. 
\subsubsection{Probes of the energy spectrum and wavefunctions}
To understand the nature of the bulk and surface states we compute the average density of states {(DOS)}, typical density of states (TDOS), and the  level statistics through the adjacent gap ratio. It is well known that the DOS %
cannot distinguish between extended and localized states whereas
the TDOS can and is a measure of the mobility gap, which can therefore be used to determine the phase boundaries. Its important to stress that 
upon the introduction of disorder the energy gap becomes soft due to non-perturbative Lifshitz states that fill in the gap. Instead, the mobility gap, or the gap in the typical density of states will distinguish metallic and insulating phases.
We therefore compute both quantities to develop a full picture of the spectrum of the model.
The average DOS is defined as
\begin{equation}\label{eq:ADOS}
\rho(E)=\frac{1}{L^2} \left[ \sum_{n}\delta(E-E_n) \right]
\end{equation}
where the sum is over all the eigenstates $|E_n\rangle$ of the Hamiltonian operator $H$ in Eq.~\eqref{eqn:fullH}, $n=1,\dots,4L^2$ 
for a linear system size $L$, $E_n$ are the eigenvalues of $H$, and $[\dots]$ denotes a disorder average. For the computation of the DOS and TDOS we average over $5000$ disorder configurations.
\begin{figure*}
\includegraphics[width=16cm]{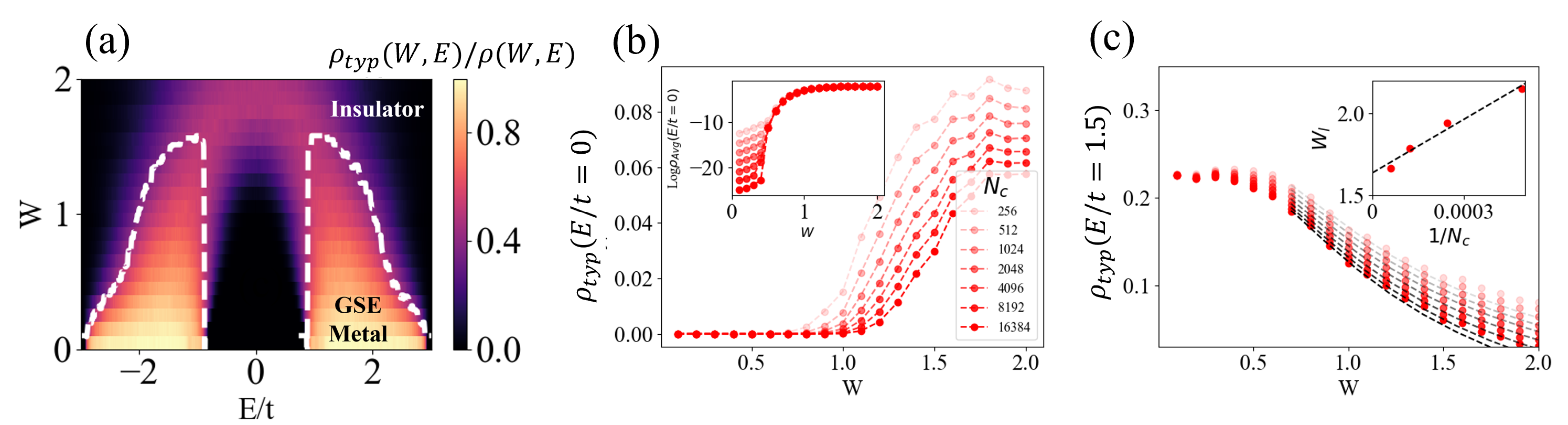}
\caption{{\bf Typical density of states in the bulk:} (a) The ratio of the bulk typical density of states (TDOS) defined in Eq.\eqref{eq:tdos} and the average DOS defined in Eq.\eqref{eq:ADOS} as a function of disorder strength, $W$ and energy $E$ for a fixed system size of $L=1000$ and KPM  expansion order $N_C=16384$. While the data at fixed $N_C$ looks like a mobility gap closes, this is not the case as the scaling with $N_C$ needs to be taken into account. The mobility edge is marked by white dashed line. It is computed following the method detailed in (c). (b) Bulk TDOS at zero energy as a function of $W$ for various expansion orders $N_{C}$ used in the KPM. (c) Bulk TDOS at $E/t=1.5$ as a function of $W$ for various expansion orders $N_{C}$ used in the KPM. The localization transition point, $W_{l}(N_{c})$, is determined by extrapolating $\rho_{typ}(E/t=1.5)$ to zero. Inset demonstrates extrapolation of $W_{l}(N_{c})$ to $N_{c}\rightarrow \infty $ limit in order to determine the phase boundary. 
}
\label{fig:BDOS}
\end{figure*}

To probe the nature of localized states we define the TDOS through the local DOS via
\begin{equation}\label{eq:tdos}
\rho_{\mathrm{typ}}(E)=\text{exp}\left(\frac{1}{4N_s}\sum_{i=1}^{N_s}\sum_{\sigma=1}^{2}\sum_{\tau=1}^{2}[\ln\rho_{i,\sigma,\tau}(E)]\right),
\end{equation} 
where  $N_s$ indicates a small collection of random sites considered $N_s \ll L^2$
and the local DOS at site $i$ for spin $\sigma$ and orbital $\tau$, $\rho_{i,\sigma,\tau}(E)$, is defined as, 
\begin{equation}\label{eq:LDOS}
    \rho_{i,\sigma,\tau}(E)=\sum_{k,\alpha,\beta}|\langle k, \alpha, \beta|i, \sigma, \tau\rangle|^2\delta(E-E_{k\alpha\beta}).
\end{equation}
When examining the surface DOS, the sum over $i$ is restricted to values on the surface of the sample. No such restriction is imposed for the bulk DOS. 
\par
In order to correctly ascertain the localization properties it is essential to reach large system sizes. Therefore, both quantities are computed using the kernel polynomial method (KPM)\cite{kpm} for a linear system size of $L=1000$, this method expands the quantity of interest in terms of Chebyshev polynomials to an order $N_C$ and is able to reach large system sizes by utilizing efficient sparse matrix-vector multiplication. The system size is sufficiently large such that  the dominant ``finite size'' effect comes from the KPM expansion order denoted $N_C$~\cite{Pixley-2016B}.
In this work we utilize the scaling of the typical DOS with the KPM expansion order ($N_C$) as a means to probe the localization transitions~\cite{kpm,Pixley-2015}. Due to the convolution of the wavefunctions and the energy spectrum, the KPM smearing of the energy is turned into a smearing of the wavefunction. In a metallic phase, the typical DOS will be $N_C$ independent, whereas in the localized phase the typical DOS will be zero in the thermodynamic limit, thus it will go to zero as $N_C\rightarrow \infty$ in our numerical simulations on finite size and expansion order. To conclude, we label the gap in the typical DOS as the mobility gap, and strictly speaking there is no gap in the average DOS for any disorder strength.

\par 
To provide an additional probe of the delocalized phase and to ascertain its diffusive (i.e. random matrix theory like) metallic properties, the level statistics is computed through the adjacent gap ratio, $r_{i}$, and averaged over 500 disorder configurations. We compute the adjacent gap ratio using exact diagonalization from, 
\begin{equation}\label{eq:LevelSpace}
    r_{i}=\frac{\text{min}(\delta_{i},\delta_{i+1})}{\text{max}(\delta_{i},\delta_{i+1})},
\end{equation}
where $\delta_{i}=E_{i+1}-E_{i}$ is the difference between neighboring, distinct eigenvalues.
In the following, the level statistics is computed as a function of energy that we average over 500 disorder configurations for linear system sizes $L=20,30,40$ that have periodic boundary conditions.
The model at hand is in the two-dimensional symplectic class (in the Cartan classification, class AII)  due to the spin-orbit coupling. Thus, upon the introduction of disorder the metallic phase is stable and can be described by random matrix theory with level statistics that satisfy the Gaussian symplectic ensemble (GSE)\cite{levelstats}. For extended states we therefore expect that $\langle r \rangle \approx0.67$, satisfying the GSE while localized states follow the Possion statistics with  $\langle r \rangle \approx 0.386$.

\subsubsection{Probes of the spin resolved spectrum and wavefunctions}
\label{sec:spin-observables}
The presence of non-trivial spin-resolved topology is determined by the spectra of the PSO ($\hat{\mathcal{S}}$), which we refer to as the spin spectrum. As we will see, the gap in the spin spectrum is as fundamental to this problem as the electronic spectrum, both must be studied in tandem. We therefore, carefully spell out our analogous approach defining relevant quantities that are similar to the energy spectrum here. 

The PSO is a function of the occupied states, fixing a given Fermi energy, $E$
{
we define the PSO without translational symmetry through
\begin{equation}
    \hat{\mathcal{S}}
=\hat{P}(E)\hat{s}\hat{P}(E)
\end{equation}
}
where $\hat{P}(E)$ projects onto all occupied states up to energy ($E$), namely $\hat{P}(E)=\sum_{E_n<E}|E_n\rangle\langle E_n|$. 
From this we define the disorder averaged spin resolved DOS at the Fermi enegy, $E$, from the spectrum of $\hat{\mathcal{S}}$ through
\begin{equation}\label{eq:SADOS}
    \rho^{\mathcal{S}}(\lambda;E)=\frac{1}{L^2}\left [\sum_{n}\delta(\lambda - \lambda_n) \right]
\end{equation}
where $\lambda_n$ denotes the eigenvalues of $\hat{\mathcal{S}}$ with eigenvectors $|\lambda_n\rangle$, and the sum is over all the eigenvalues $i=1,\dots,4L^2$.

However, its not sufficient to know if the spin spectrum is finite we have to also determine if the spin eigenvectors are localized in real space or not. Therefore,
analogous to the local DOS we define the spin resolved local DOS at site $i$ for spin $\sigma$ and orbital $\tau$ as
\begin{equation}
     \rho_{i,\sigma,\tau}^{\mathcal{S}}(\lambda;E)=\sum_{k,\alpha,\beta}|\langle k,\alpha,\beta|i,\sigma,\tau \rangle|^2 \delta(\lambda-\lambda_{k\alpha\beta}),
\end{equation}
from which we define the  typical spin resolved DOS
\begin{equation}\label{eq:SpinTDOS}
    \rho_{\mathrm{typ}}^{\mathcal{S}}(\lambda;E)=\text{exp}\left(\frac{1}{4N_s}\sum_{i=1}^{N_s}\sum_{\sigma=1}^{2}\sum_{\tau=1}^{2}[\ln\rho_{i,\sigma,\tau}^{\mathcal{S}}(\lambda;E)]\right).
\end{equation}
Similar to a metallic and insulating TDOS, we expect that whether or not the spin resolved TDOS is non-zero  will tell us if the eigenmodes of the PSO are localized or not. Similarly, the gap in the spin resolved TDOS, we refer to as the spin mobility gap.

In the following we study the spin resolved spectrum using exact diagonalization and consider system sizes up to $L=20$. 
{Despite the small sizes we are able to get rather conclusive data on the nature of the PSO in the presence of disoder, which for the present problem has much less severe finite size effects compared to the eigenstates and spectrum of the Hamiltonian. }
We remark that it is possible to construct a KPM description of the spin spectrum to reach larger sizes but it requires a double Chebyshev expansion and is therefore left for future development. 

\subsubsection{Topological properties}\label{sec:TopoProperties}
\par 
In the presence of finite disorder we again diagnose bulk topology through computation of the spin-resolved Chern number via a modifying of the coupling matrix method described in Ref. \cite{zhang2013coupling}. We emphasize that our implementation differs only through a replacement of the real-space Hamiltonian with the PSO. For clarity, we briefly desctibe the method.
\par 
Defining twisted boundary conditions for single-particle eigenfunctions of the PSO as $\psi(x+L,y)=e^{i\theta_{x}}\psi(x,y)$ and $\psi(x,y+L)=e^{i\theta_{y}}\psi(x,y)$ such that $\theta_{i}\in [0,2\pi)$. The single-particle eigenfunctions can be Fourier transformed. The twisted boundary conditions impose the constraint that reciprocal space coordinates take on discrete values, $\mathbf{k}=\left(\frac{2n_{1}\pi}{L},\frac{2n_{2}\pi}{L}\right)+\left(\frac{\theta_{x}}{L},\frac{\theta_{y}}{L}\right)$, where $0\leq n_{i}\leq L$.
\par 
The Chern number is then computed as, 
\begin{equation}\label{eq:SpinChernN}
    C_{\downarrow}=\frac{1}{2\pi i}\oint_{\partial R_{q}}d\mathbf{l}_{q}\langle  \Phi_{q}|\nabla_{q} \Phi_{q}\rangle,
\end{equation}
where $|\Phi_{q}\rangle$ is the Fourier transform of the negative eigenstates of the PSO and $\partial R_{q}$ is the boundary of the square of side length $2\pi/L$. To carry out the computations we discretize the $\partial R_{q}$ into eighty segments, replacing the derivative with finite differences. Finally, we average over a minimum of 20 disorder configurations.

\begin{figure}
    \centering
    \includegraphics[width=8cm]{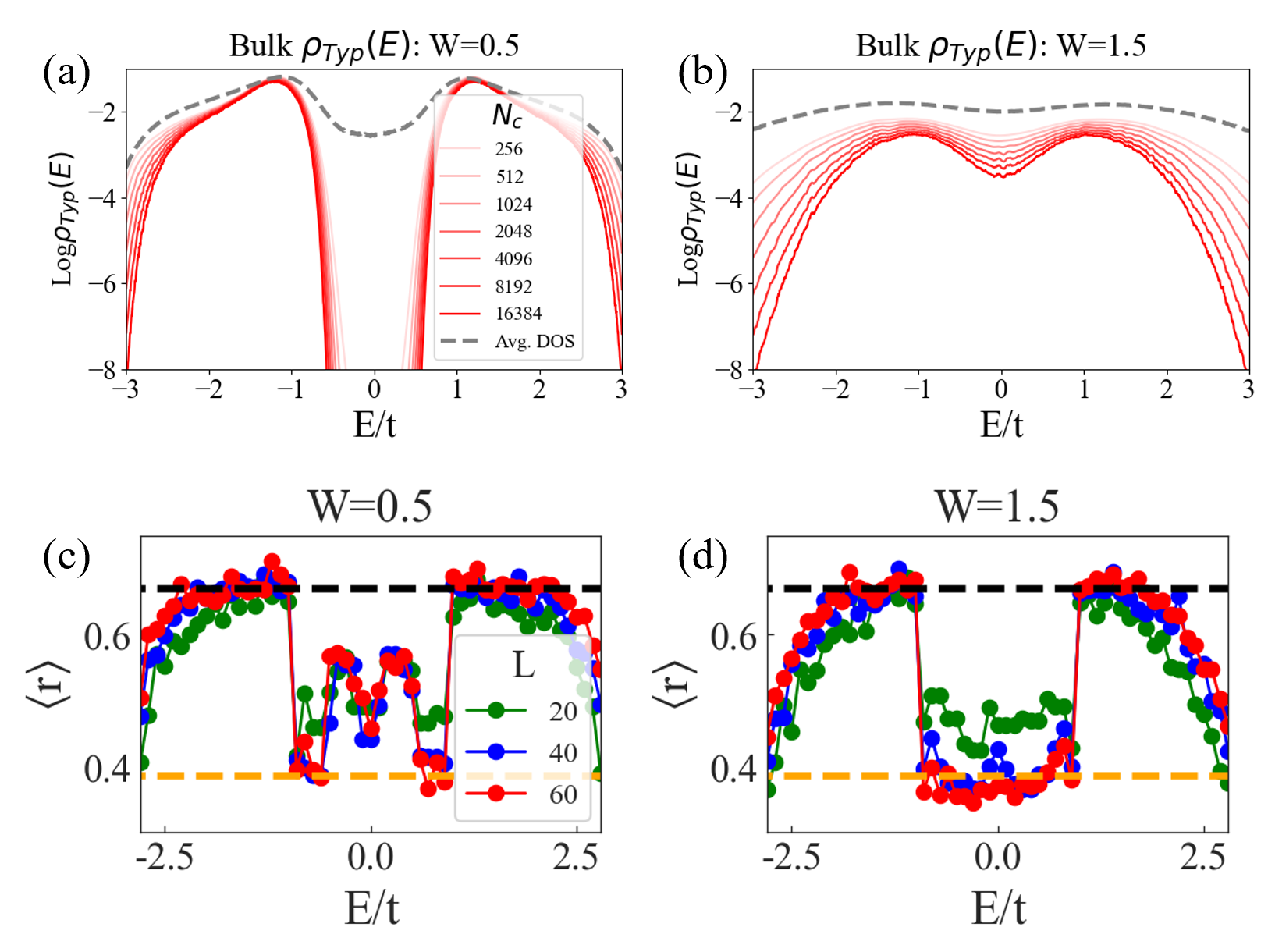}
    \caption{{\bf Revealing the GSE metal at finite energy}: Bulk TDOS define in Eq. \eqref{eq:tdos} as a function energy, varying the expansion order, $N_{C}$ for the kernel polynomial method, fixing the disorder strength to (a) $W=0.5$ and (b) $W=1.5$. (c)-(d) Adjacent gap ratio defined in Eq. \eqref{eq:LevelSpace} as a function energy for a linear system size $L$ with periodic boundary conditions, fixing the disorder strength (c) $W=0.5$ and (d) $W=1.5$ and averaging over 500 disorder configurations. For increasing disorder strength the region supporting finite energy states that obey the expected result for a GSE, $\langle r \rangle \approx 0.6750$ (marked with a black dashed line) becomes sharper. In contrast, in the localized regimes [where we see a strong $N_C$ dependence in (a) and (b)] we also see the level statistics are Poisson (or approaching it) with $\langle r \rangle = 2\log 2-1 \approx 0.39$ (marked by dashed orange line) showing the two results are nicely compatible.}
\label{fig:BDOS2}
\end{figure}

\begin{figure*}
\includegraphics[width=16cm]{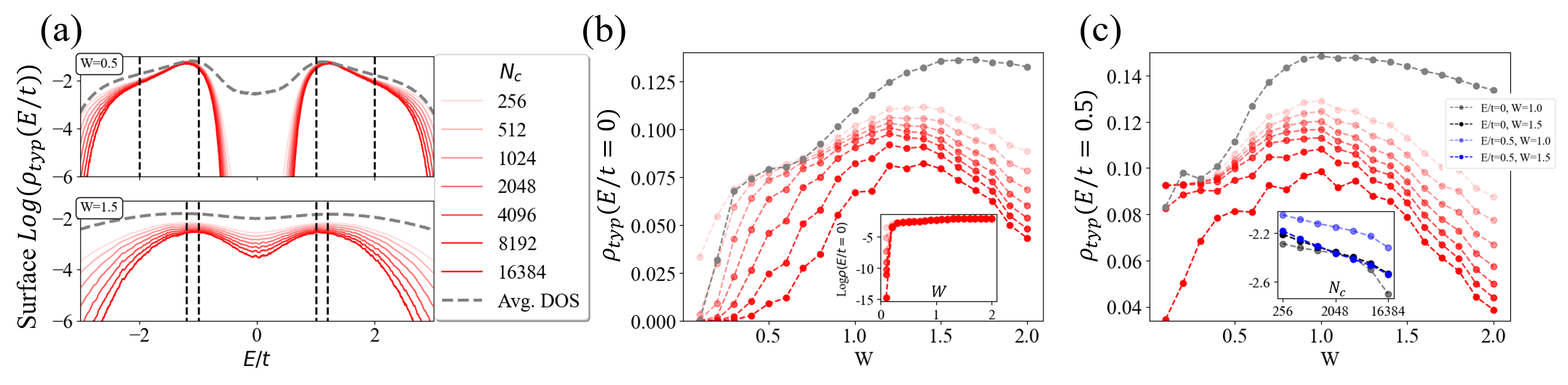}
\caption{{\bf TDOS computed for sites on an open $x$ surface}: We show the TDOS as defined in Eq. \eqref{eq:tdos} as a function energy, varying the expansion order, $N_{C}$ for the kernel polynomial method, and fixing the disorder strength to (top) $W=0.5$ and (bottom) $W=1.5$. Black dashed lines approximate the location of the bulk mobility gap. (b) Dependence of surface $\rho_{\mathrm{typ}}$ as a function of expansion order, $N_{C}$, at zero energy. Inset in shows convergence of average DOS at zero energy. (c) Dependence of surface $\rho_{\mathrm{typ}}$ as a function of expansion order, $N_{C}$, at finite energy. Inset displays $N_{c}$ dependence at varying energy and disorder strength. 
}
\label{fig:SDOS}
\end{figure*}

\section{Effects of Disorder}\label{sec:disorder}
\par 

We now come to the effects of disorder when starting from a  HOTI band structure in two-dimensions. Importantly, the disorder breaks the translational symmetry and the identification of each phase then follows from the quantities defined in Sec.~\ref{sec:observables}.

\subsection{Bulk and Surface Electronic Phase Diagram}\label{sec:ElectronicDisorder}
\par

We start by establishing the nature of the bulk phase of the model as a function of disorder.
The TDOS as a function of disorder strength and energy is shown in Fig.~\eqref{fig:BDOS}(a). 
Importantly, we see a clear mobility gap at weak disorder that does not close as we increase the disorder strength. Instead, we see two clearly metallic bands at finite energies, which Anderson localize around a disorder strength of $W=1.5$. To mark the mobility edge we follow the procedure outlined in Ref. \cite{Pixley2016RareRegion}, extrapolating $\rho_{typ}(E/t)$ to to zero as a function of $W$ for each KPM expansion parameter $N_{c}$, identifying the localization transition, $W_{l}(N_{c})$. A second extrapolation is performed to obtain $W_{l}(N_{c})$ in the limit $N_{c}\rightarrow \infty$. Repeating this procedure at each value of $E/t$ produces the phase boundary shown as a dashed white line in Fig.~\eqref{fig:BDOS}(a).

The absence of the closure of the mobility gap  is shown clearly through the $N_C$ dependence of $\rho_{\mathrm{typ}}(E=0)$ at the center of the band in Fig.~\eqref{fig:BDOS}(b), where the typical DOS always is a decreasing function of increasing $N_C$ suggesting it is always insulating.
However, the average DOS is converged in $N_C$ (on this linear scale)  and is always non-zero at finite $W$, though exponentially small at weak disorder. This gives it an artificial sense of looking like it is lifting off from zero at a particular disorder strength on a linear scale, but in facts its just exponentially small in the disorder strength as expected for Lifshitz states. This is in strong contrast to the localizing behaviour seen in the typical DOS. %

In contrast, sitting at a finite Fermi energy $E/t=1.5$, deep in the metallic band, we see a stable metallic phase, which develops a strong $N_C$ dependence at large disorder strength. This is displayed in Fig.~\eqref{fig:BDOS}(c), where the power law fit to $\rho_{Typ}(E)$ as a function of {$\sim(W-W_l(N_C))^{x}$} to determine an estimate of the localization transition at this expansion order denoted $W_{l}(N_C)$, which is plotted as black dashed lines. In our data, we find optimal fits for $1.4<x<2.0$.
The inset then further shows the extrapolation of $W_{l}(N_C)$ to $N_{c}\rightarrow \infty$ to obtain the result in the thermodynamic limit. Taken together, the results in Fig. \eqref{fig:BDOS} clearly show two metallic bands undergoing an Anderson localization transition. If we take a Fermi energy in the mobility gap center we see that the localization transition takes place {\it below the Fermi energy}. This phenomena is quite unlike that of a Chern  or $\mathbb{Z}_2$ topological insulator where the mobility gap closes in the band center at a critical disorder strength, namely the delocalization-localization transition  takes place at the Fermi energy. Here, for the case of the HOTI we find that this is no longer the case. 

%
%
 To describe the nature of the metallic phase  we turn to the average level statistics on small sizes and the typical DOS at much larger sizes to provide a comparison in Fig.~\ref{fig:BDOS2}. Importantly, we find that the electronic bands have an 
 adjacent gap ratio consistent with the GSE, thus demonstrating a delocalized phase in two-dimensions. Importantly, the conclusions of the scaling of the TDOS with $N_C$ agrees well with the level statistics for estimate of the region of the metallic phase.

\begin{figure}
\includegraphics[width=8cm]{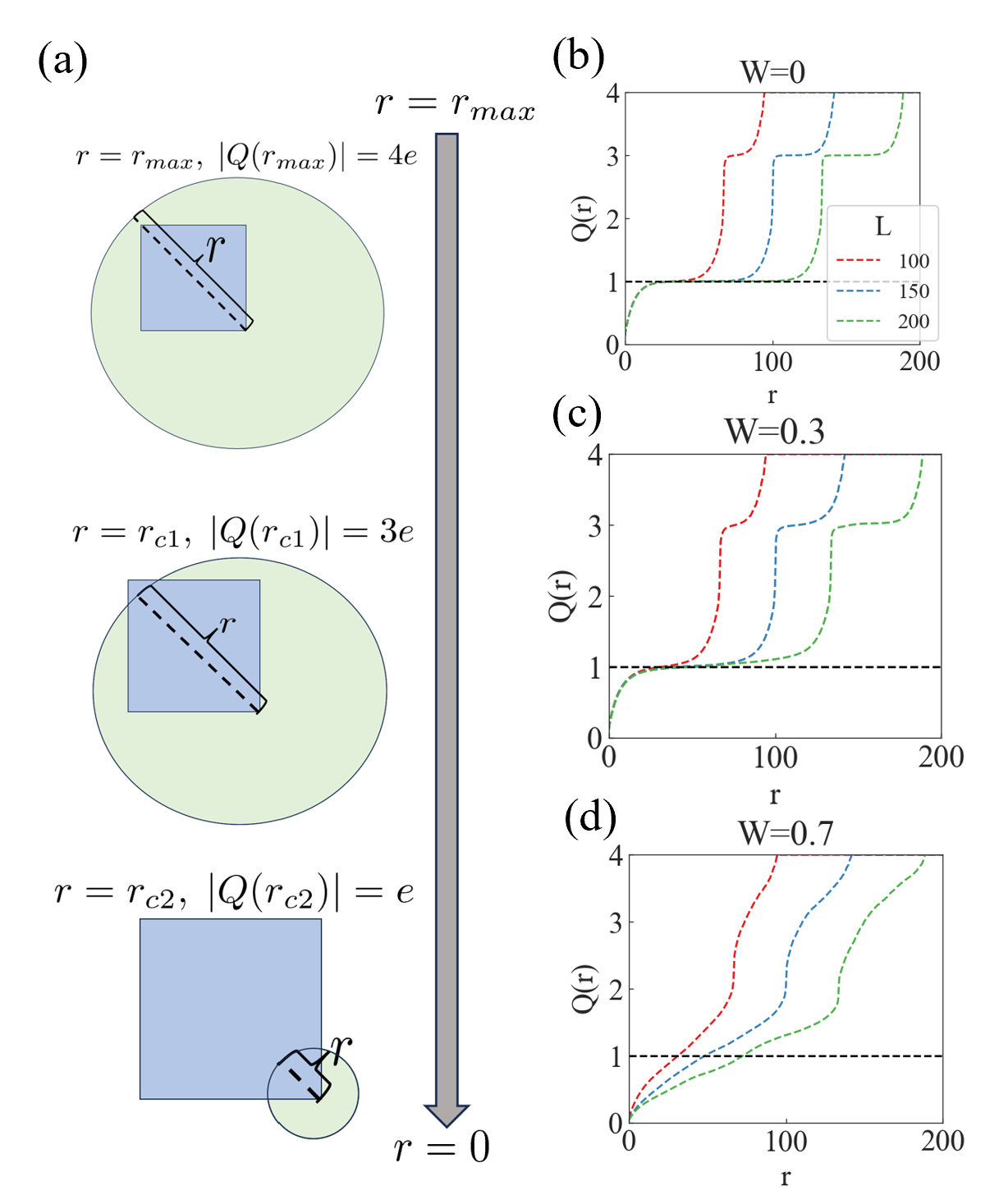}
\caption{{\bf Delocalization of the corner charge}: (a) Schematic detailing measurement of corner bound charge. Charge density within a circle of radius, $r$, centered at a corner is computed. $Q(r)$, defined in Eq. \eqref{eq:Cornercharge}, exhibits three main features: If $r$ is large enough to contain the full system, $Q(r)=4e$, if $r$ is decreased such that only three corners are contained within the circle, $Q(r)=3e$, two corners of the square are then removed simultaneously providing a third transition to $Q(r)=e$. (b)-(d) Charge contained with a circle of radius $r$ centered about a corner of a two-dimensional slab of size $L^{2}$ with open-boundary conditions. We consider the four lowest lying states as a function of disorder strength, averaging over 100 disorder configurations. Quantization of corner charge is destroyed by increasing of the disorder strength, indicating that corner localized modes are not protected from disorder.
} 
\label{fig:CornerCharge}
\end{figure}

\subsubsection{Nature of edge modes at finite energy}\label{sec:Surface}
To complete our analysis, we determine the phase diagram of the surface states. To reveal the edge states at finite energy we impose periodic boundary conditions along the $y$ direction and open boundary conditions along the $x$ direction. We study the surface TDOS  by modifying Eq.~\eqref{eq:tdos} to restrict the sum over $i$ to lattice sites on the $x$ edge. In Fig. \eqref{fig:SDOS}(a) we plot the surface TDOS as a function of energy at representative values of disorder strength. We do not find any regime that displays stable metallic surface states and we observe a clear lack of convergence for increasing $N_{C}$, that demonstrates the surface states are localized at all energy. To clarify this behavior, we plot the TDOS at $E/t=0.5$ and $E/t=0$ as a function of disorder strength in Fig. \eqref{fig:SDOS} (b) and Fig. \eqref{fig:SDOS}(c) respectively. At zero energy, the TDOS indicates $\rho_{\mathrm{typ}}(E/t=0)\rightarrow 0$ as $N_{C}\rightarrow \infty$ is always true. By contrast, the ADOS is non-zero at all values of finite disorder, demonstrating the presence of disorder induced, localized zero energy states on the surface. These zero-energy states have been the source of speculation that a higher-order to first-order transition can occur. Our analysis unambiguously shows that these states are \emph{not} topologically protected and no such transition has occurred.  In summary, our numerical data implies that the surface states localize for infinitesimal disorder strength.

\begin{figure*}

    \includegraphics[width=16cm]{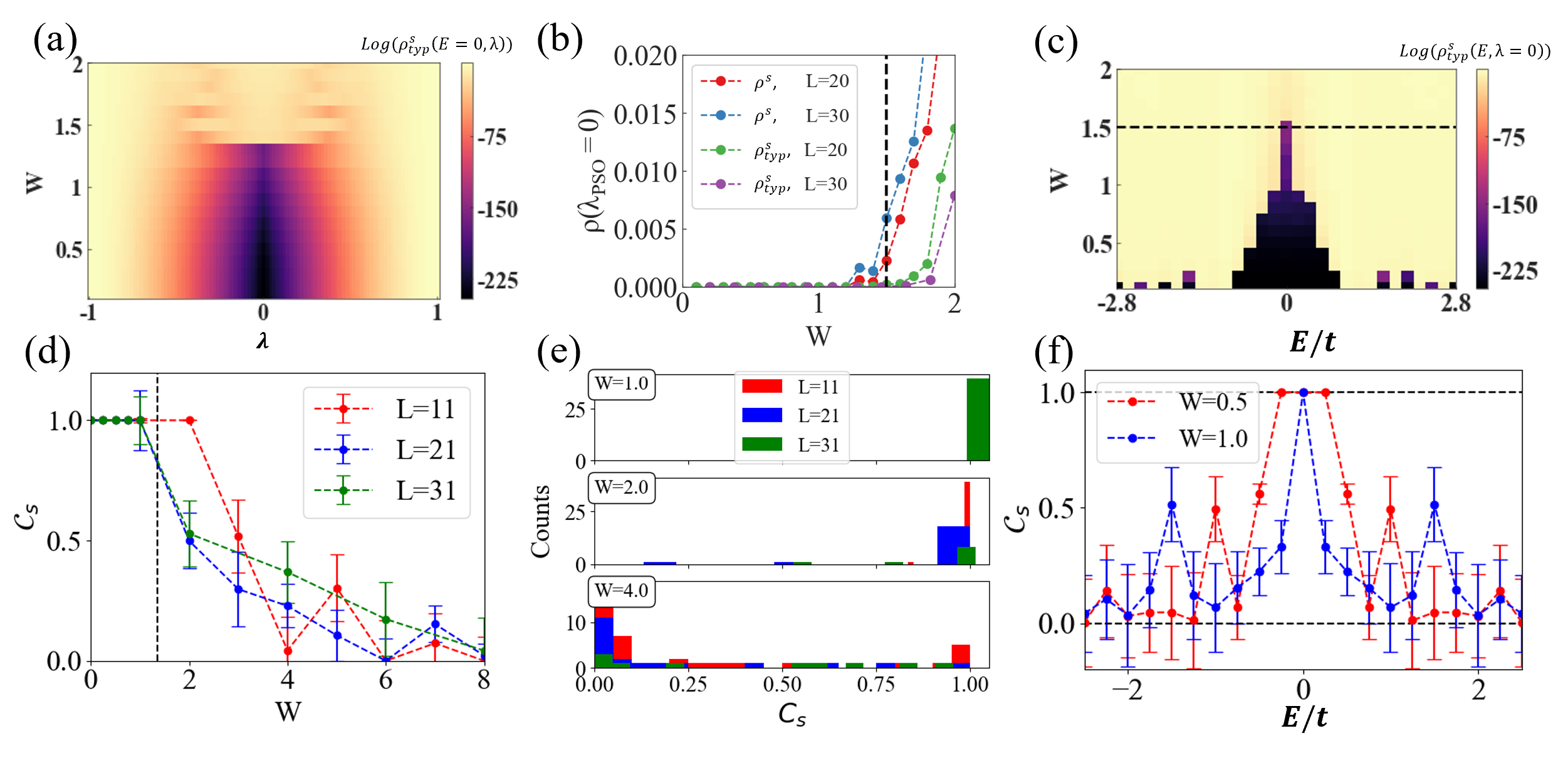}
    \caption{
    {\bf Disordered properties of the PSO and spin Chern number}:
    (a) Typical density of states of projected spin-operator (PSO), defined in Eq. \eqref{eq:SpinTDOS}, at $E=0$ as a function of magnitude of eigenvalues for the PSO, $(\lambda_{PSO})$, and disorder strength under periodic boundary conditions. (b) Typical and average density of states, defined in Eq. \eqref{eq:ADOS}, of projected spin-operator (PSO) defined at $E=0$ as a function of disorder strength at $\lambda_{PSO}=0$.  (c) Typical density of states of projected spin-operator (PSO) as a function of disorder strength and Fermi energy at $\lambda_{PSO}=0$, which shows that it becomes finite wherever the spin-gap closes. In each plot a dashed line marks the approximate location at which all occupied states localize in the bulk. (d) Calculation of spin-Chern number, defined in Eq.~\eqref{eq:SpinChernN}, at the Fermi energy for a supercell of linear size $L$ as a function of disorder strength, averaging over 20 disorder configurations. (e) Histogram detailing results of spin-Chern number computation for individual disorder configurations as a function of disorder strength, fixing $E=0$. (f) Spin-Chern number as a function of the Fermi energy for representative values of disorder strength.
    }
    \label{fig:SpinResDisorder}
\end{figure*}

\subsubsection{Nature of the corner modes}\label{sec:C4Corner}

The robustness of corner modes to the introduction of disorder in two-dimensional HOTIs has been studied for alternate models~\cite{PhysRevB.99.085406}. Such studies are broadly interested in maintaining  the bulk-corner correspondence of HOTIs as dictated by topological invariants protected by symmetries that may be removed through the inclusion of disorder. In the case of this work, the bulk invariant protecting the corner modes is a quantized quadrupole moment, $Q_{xy}$. As stated in Sec.~\eqref{sec:Clean}, quantization of the quadrupole moment is protected by chiral symmetry, in the clean limit. Introduction of onsite disorder removes the chiral symmetry. It is therefore expected that corner modes will not be protected. 
\par 
To investigate the fate of the corner modes upon introduction of disorder, using Lanczos\cite{lanczos1950iteration} we consider the four-lowest lying states of a two-dimensional slab of size $L^{2}$ at $L=100,150,200$ with open boundary conditions along both the $x$ and $y$ directions. The charge localized at the corner is computed as, 
\begin{equation}\label{eq:Cornercharge}
    Q(r)/e=\sum_{\mathbf{r'}_{i}<r}\sum_{n=1}^{4}|\psi_{n}(\mathbf{r'}_i)|^2 ,
\end{equation}
where the sums are over the four lowest lying state and $\mathbf{r'}_{i}$ indicates distance from the bottom right corner of the two-dimensional slab. A schematic of the expected behavior in the clean limit is shown in Fig. \eqref{fig:CornerCharge}(a). 
\par
The results in Fig. \eqref{fig:CornerCharge}(b)-(d) demonstrate that in the presence of disorder charge quantization at the corners is destroyed. This is in accordance with the removal of the underlying symmetry protected topology. 

\subsection{Bulk and Surface Spin Resolved Phase Diagram and Topology}\label{sec:SpinDisorder}
\par 
Suprisingly, our analysis of the electronic spectrum indicates no extended states are observed at zero energy regardless of the disorder strength. When studying the evolution of a topological insulator as a function of disorder, it is common to focus on the properties of the system at zero energy due to the understanding that the non-trivial topology is protected by the bulk-mobility gap and upon closing the gap,  the wavefunctions delocalize. The model at hand exhibits a strikingly different behavior. Increasing the magnitude of disorder drives a series of metal-insulator transitions at finite-energy away from the band center (as shown in Fig.~\ref{fig:BDOS}). However, in the clean limit in the presence of a finite spin Chern number it is necessary to study both the electronic and spin gap. Therefore, we have generalized each of these theoretical probes from the energy to the spin spectrum in Sec.~\ref{sec:spin-observables} and in this section we turn to a study of the properties of the PSO in the presence of disorder.
\par 
To do so we first fix the Fermi energy $E=0$ and compute the TDOS of the PSO for a system size of $L=20$ under PBCs following Eq.~\eqref{eq:SpinTDOS}, averaging over 50 disorder configurations. The results in Fig.~\eqref{fig:SpinResDisorder}(a) are distinct from what was observed for the electronic spectrum in Fig.~\ref{fig:BDOS}. The spin spectrum supports a spectral gap that narrows with increasing disorder strength and vanishes at $W\approx 1.5$ as shown in Fig. \eqref{fig:SpinResDisorder}(b). Interestingly, this value approximately aligns with the localization of all bulk states. 
\par 
Next, we consider the effects of varying $E$, computing the TDOS of the PSO at $\lambda_{PSO}=0$ to determine the presence/absence of a spin-gap as a function of varying the Fermi energy and disorder strength. The results shown in Fig.~\eqref{fig:SpinResDisorder}(c), demonstrate that as disorder strength is increased, the range of $E$ for which a spin-gap exists begins to narrow, reflecting the behavior of a familiar Chern insulator. 
Thus, we have confirmed that the spin mobility gap closes at this topological transition.
\par
\subsubsection{Topology of the spin spectrum}
The spin-Chern number is expected to remain robust as long as the electronic and spin mobility gaps remain non-zero. It is further reasonable to expect that the spin-Chern number is trivialized upon localization of all occupied bulk states in the electronic spectrum. To verify these expectations, we first compute the bulk spin-Chern number fixing $E=0$.  The results shown in Fig.~\eqref{fig:SpinResDisorder}(d) demonstrate that as we increase the linear system size, $L$, the spin-Chern number is quantized for weak disorder, deviating from the quantized value for $W\gtrsim 1.5$. 
{This is in good agreement with the analysis of both the TDOS for the spin and electronic spectrum, demonstrating that the closure of the spin spectral gap and localization of all occupied states are both realized at the same disorder strength (to within our numerical accuracy). } To provide further evidence of a sharply defined spin Chern number and the nature of its fluctuations we show the distribution of the spin Chern number across different disorder samples in Fig.~\eqref{fig:SpinResDisorder}(e). This reveals that the spin Chern number is quantized for every sample not just on average. As we go through the transition and increase disorder strength this distribution develops a tail towards zero.
\par
Last, we further verify the correspondence between the presence of a mobility gap in the spin-spectrum and a quantized spin-Chern number we compute the spin-Chern number as a function of $E$ at representative values of the disorder strength $W=0.5$ and $W=1.0$ in Fig. \eqref{fig:SpinResDisorder}(f). The results demonstrate
that the spin-Chern number remains quantized when the Fermi Energy ($E$) is in the electronic and spin mobility gaps.
\par 
Having firmly established a direct correspondence between the existence of a spin-gap and the existence of a quantized spin-Chern number, a schematic of the bulk phase diagram is constructed. This requires knowledge of the bulk-mobility gap as computed in Fig. \eqref{fig:BDOS}, as well as knowledge of the spin mobility gap as computed in Fig.~\eqref{fig:SpinResDisorder}. The resulting phase diagram, m shown in Fig. \eqref{fig:c4phase}, is striking as it implies the possibility of a topological phase transition, both as a function of the Fermi energy and disorder strength, %
without closing the mobility gap.
However the PSO, which controls the bulk topology, does display such conventional gap closing as seen in Fig. \eqref{fig:SpinResDisorder}(c). It is the hidden nature of the PSO and the interplay between the bulk mobility gap and spin-gap which allows for the construction of such an exotic phase diagram. 
\par 

\subsubsection{Spin resolved edge modes}
\par

\begin{figure}
    \centering
    \includegraphics[width=8cm]{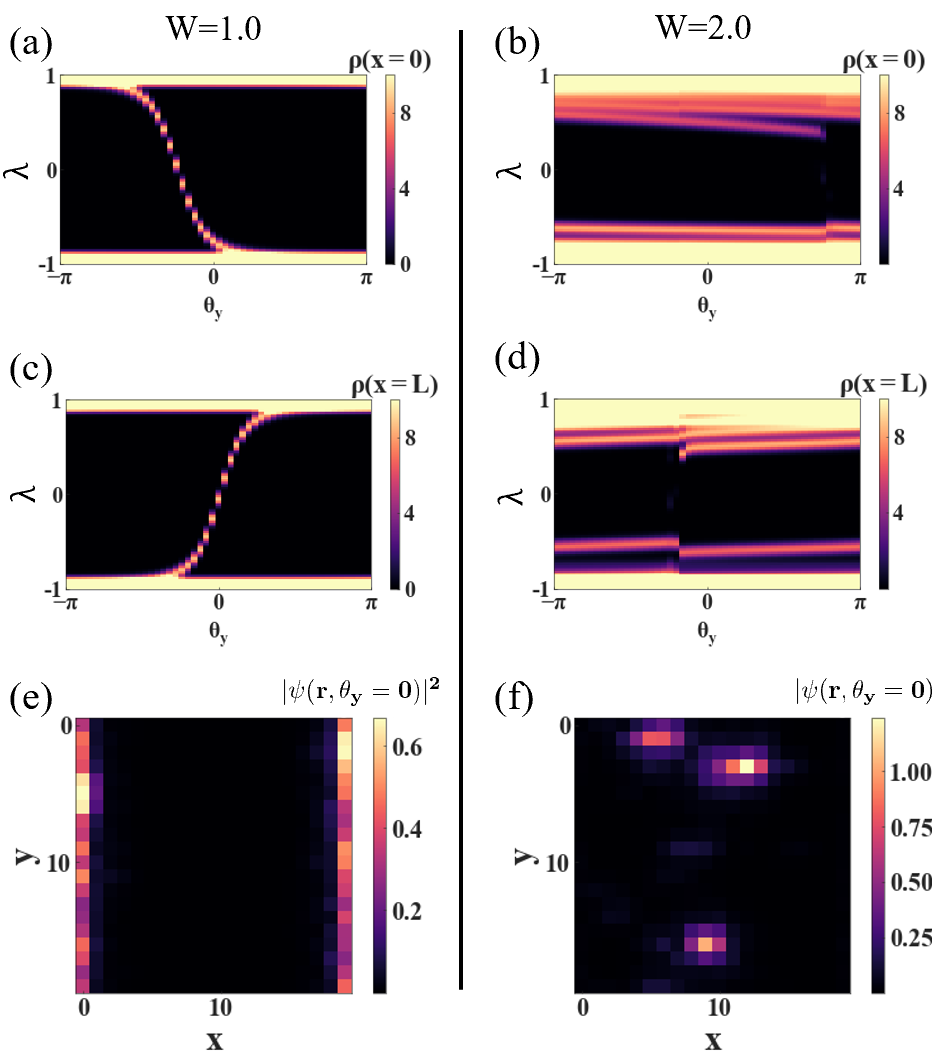}
    \caption{{\bf Surface states of the PSO}: The PSO on an $L=20$ size system with open boundary conditions along $x$ and twisted boundary conditions along $y$, for twist $\theta_{y}$, is considered. The average local density of states of the PSO, defined in Eq. \eqref{eq:SADOS}, on the $x=0$ surfaces as a function of the twist angle are shown for $W=1.0$ in (a) and $W=2.0$ in (b). The average local density of states of the PSO on the $x=L$ surfaces as a function of the twist angle are shown for $W=1.0$ in (c) and $W=2.0$ in (c).  The results demonstrate the presence(absence) of topological gapless edge modes when a spectral gap is present(absent) in the PSO. The real-space distribution of the eigenstates nearest to $\lambda=0$ for twist angle $\theta_{y}=0$ are given in (e) and (f) for $W=1$ and $W=2$ respectively.} 
    \label{fig:PSOEdge}
\end{figure}

In Sec. \eqref{sec:Surface} our computation of the surface TDOS on the edge in a cylindrical geometry led to the conclusion that topological edge states are all localized at any $W>0$. Nevertheless, it is interesting to investigate the ``surface spectra" of the PSO for a fixed $E$. If we consider the eigenstates associated with the negative (positive) eigenvalues of the PSO to be ``occupied" and "unoccupied", the magnitude of the spin-Chern number corresponds to the magnitude of the Chern number for the ``occupied states" of the PSO. As such, for a non-zero quantized spin-Chern number we expect the PSO to support chiral edge modes, a hall mark of the Chern insulator state. We directly examine these edge modes for a system size $L=20$ in a cylindrical geometry (open boundary conditions along $x$ and twisted boundary conditions along $y$). We consider representative disorder strengths $W=1$ and $W=2$ for which the spin gap has been shown to be open and closed, respectively. 

\par 
In order to examine the edge modes, we track the local density of states along the $x=0$ and $x=L$ edges as a function of varying the twisted boundary conditions along $y$. The results, displayed in Fig. \eqref{fig:PSOEdge}(a) and (c) for $W=1$ and Fig. \eqref{fig:PSOEdge}(b) and (d) for $W=2$ respectively are in accordance with classification of the PSO as a Chern insulator. For $W=2$ the spectral flow vanishes indicating the trivial topology. The real-space distribution of the eigenstates nearest to $\lambda=0$ for $\theta_{y}=0$ are also shown in Fig. \eqref{fig:PSOEdge}(e) and (f) for $W=1$ and $W=2$ respectively. These figures underscore that in the topological phase the states are bound to the edge and decouple from the edge in the trivial phase. While the results in Fig. \eqref{fig:PSOEdge} are for single disorder configurations, we have considered 20 possible disorder configurations for each value of $W$, confirming the results are consistent in each case.

\subsection{Signatures of spin-resolved topology in the electronic spectrum}\label{sec:flux}
\par
The previous sections demonstrate that the spin-resolved topological Anderson insulator phase does not admit protected boundary modes at a surface of co-dimension $n\neq 0$. It is then natural to investigate the case of $n=0$ through introduction of a zero-dimensional defect in the lattice. The defect we consider is that generated by a dislocation or equivalently insertion of a magnetic flux tube. This is a natural choice as flux insertion is a well-established probe of both Chern insulators and quantum spin-Hall insulators and should allow for the topology of the spin-spectrum to be probed through an analysis of the electronic spectrum \cite{QiSpinCharge,SpinChargeVishwanath,slager2012,MESAROS2013977,juricicprobe,schindler2022topological,tynerbismuthene,Lin2022Spin}.
\par
\begin{figure}
\includegraphics[width=8cm]{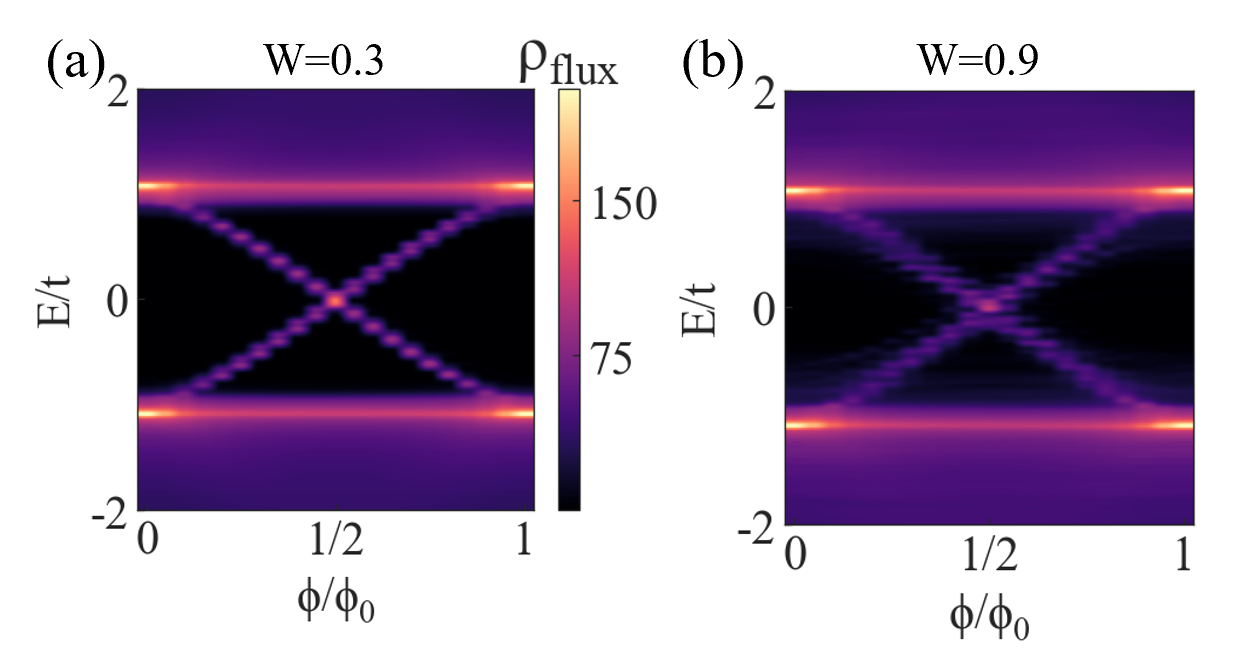}
\caption{{\bf Probing the topology with a magnetic flux tube}: Local density of states on inserted flux tube ($\rho_{flux}$) as a function of flux strength fixing (a) $W=0.3$ and (b) $W=0.9$, considering a linear system size $L=100$ and averaging over $100$ disorder configurations. At $\phi=\phi_{0}/2$ the modes are degenerate and act in an identical manner to end states of a spinful SSH chain.
}
\label{fig:BulkTopo}
\end{figure}

The magnetic flux tube is placed at the center of the disordered, real-space Hamiltonian of linear system size $L$, centered at the origin. The presence of a flux tube is simulated using the Peierls substitution with the choice of gauge connection given by 
\begin{equation}
 A_{y}= \phi \delta(y) \Theta(x).    
\end{equation}
Imposing a cylindrical geometry we calculate the disorder averaged density of states on the defect, varying the flux strength $\phi$ from $0$ to the flux quanta, $\phi_{0}=hc/e$. The results for a linear system size $L=100$ are shown in Fig. \eqref{fig:BulkTopo}(a)-(b) for $W=0.3$ and $W=0.9$ respectively. The flux tube binds two states which are pumped across the bulk gap as a function of the flux strength. 

To understand this behavior, consider first the case where we set $t_{1}=0$ [that is defined in $H_0$ below Eq.~\eqref{eqn:fullH}] such that spin-rotation symmetry is preserved. In this case the occupied subspace for $E=0$ is composed of the occupied subspace of two Chern insulators with $C=\pm 1$. As detailed in Ref. \cite{QiSpinCharge}, for each Chern insulator the {adiabatic} flux threading process causes $N=|C|$ modes bound to the flux tube to be pumped across the bulk gap. The sign of the Chern number further determines whether these bound modes are pumped from the occupied to unoccupied subspace or vice versa. In the case of a quantum spin-Hall insulator, we therefore observe two counter-propagating flux tube bound modes as a function of flux strength. 
\par
Upon introducing a finite spin-orbit coupling, removing the spin-rotation symmetry, it is possible to gap the degeneracy of the flux tube bound modes seen at $\phi=\phi_{0}/2$. In the clean limit, for $t_{1}\neq 0$, this degeneracy is protected by crystalline symmetry\cite{tyner2020topology,Lin2022Spin}. For finite disorder, this degeneracy is removed for individual disorder configurations but the states remain bound to the flux tube. When considering a large number of disorder configurations the symmetries are restored in the average\cite{ASPT} and the degeneracy reappears. In brief, the flux tube can be used to probe whether the occupied states contain hidden subspaces corresponding to occupied Chern insulating ground states. 
\par 

\section{Discussion and Conclusion}\label{sec:Summary}
In this manuscript we have investigated the effects of quenched short range potential disorder on the zero temperature phase diagram of two-dimensional higher order topological insulators. As the higher order topology arises from a combination of crystalline and time reversal symmetry, and both the bulk and the surface are gapped it has remained unclear what properties remain in the presence of disorder, which breaks translational symmetry in each sample. 
Our work definitively shows that the higher order topology is ``stripped away'' through localization leaving only the  the spin resolved topology that is otherwise hidden from the energy spectrum. As a consequence, future theoretical studies of disordered higher order topology must incorporate the study of the spin resolved spectrum in tandem with the energy spectrum that is conventionally analyzed. It will be exciting in future work to determine the low energy effective field theoretical description for these observations.

We expect these results to be relevant in the ongoing studies of two-dimensional higher-order topological insulators~\cite{sodequist2022abundance,costa2021discovery,Lin2022Spin,tyner2023solitons}. The outcome of the present study shows that edge modes in two-dimensional higher order topological materials are fragile and will be localized in any realistic sample.
Furthermore, our results are directly applicable in the context of two-dimensional heterostructures and twisted (or moir\'e) materials where additional degrees of freedom such as valley and mini-valley, etc. can give rise to pseudo(spin)-resolved topology~\cite{mak2014valley,tao2024giant}. It will be exciting to see how these conclusions play out when considering three-dimensional higher order topology, which we leave for future work.

Our results expand our fundamental understanding of topological materials and the meaning  of ``topological protection" in higher order topology that requires crystalline symmetries. By decoupling the spin and bulk spectral gap through disorder, the model yields a novel phase diagram whereby a topological phase is bordered by a trivial insulating phase without an intervening gapless point. 
In other words, the bulk mobility gap and the spin mobility gap have decoupled resulting in Anderson localization taking place below the Fermi energy where the topology remains quantized.
Furthermore, our results extend the concept of $n$-th order insulators in a direction not previously explored, demonstrating that topologically protected boundary modes exist only on a zero-dimensional defect. 

\acknowledgements{}
We thank Jennifer Cano, Sankar Das Sarma, Taylor Hughes, David Huse, Jay Sau, David Vanderbilt, and Justin Wilson for insightful discussions.
This work is partially supported by NSF Career Grant No.~DMR- 1941569 and the Alfred P.~Sloan Foundation through a Sloan Research Fellowship (C.G., J.H.P.). Part of this work was performed in part at the Aspen Center for Physics, which is supported by the National Science Foundation Grant No.~PHY-2210452 (A.T., J.H.P.) as well as the Kavli Institute of Theoretical Physics that is supported in part by the National Science Foundation under Grants No.~NSF PHY-1748958 and PHY-2309135 (J.H.P.).
A.T. acknowledges the hospitality of the Center for Materials Theory at Rutgers University.
Nordita is supported in part by NordForsk. A portion of this research was done using services provided by the OSG Consortium \cite{osga,osgb,osg,osg2}, which is supported by the National Science Foundation awards No. 2030508 and No. 1836650. The authors acknowledge the Office of Advanced Research Computing (OARC) at Rutgers, The State University of New Jersey for providing access to the Amarel cluster and associated research computing resources that have contributed to the results reported here. 
\bibliography{ref.bib}

\appendix
\section{System size dependence of the phase diagram}

\par
\begin{figure}[H]
    \includegraphics[width=7.5cm]{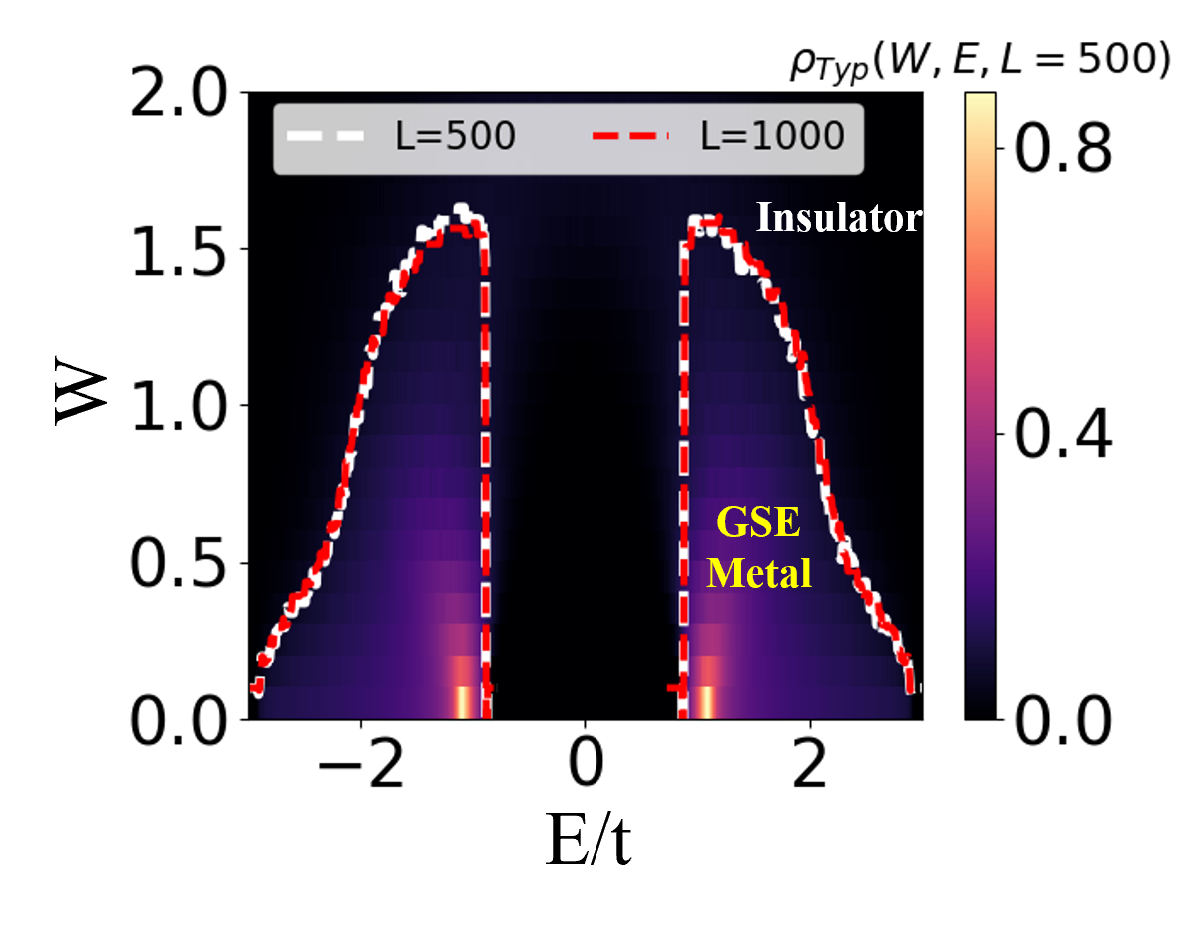}
    \caption{{\bf Typical density of states in the bulk for L=500.} The bulk typical density of states (TDOS) as a function of disorder strength, $W$ and energy $E$ for a fixed system size of $L=500$ and KPM  expansion order $N_C=16384$. The mobility edge computed for $L=500$ and $L=1000$ are marked by white and red dashed lines respectively.}
    \label{fig:App1Fig1}
\end{figure}

 
In order to verify that the main finite-size effects in the main body arise due to the KPM expansion order, $N_{c}$, and not from the physical system size, $L$, we consider a system smaller than that used to obtain the results in the main body. Namely, we consider a system of size $L=500$, the main body utilizes $L=1000$. The bulk TDOS is then recomputed as a function of varying the Fermi energy, $E$, as well as the disorder strength,$W$ and the KPM expansion order. We follow the procedure detailed in the main body to systematically obtain the phase boundary. The results in Fig. \eqref{fig:App1Fig1}, compare the phase boundary as determined for $L=500$ with that determined for $L=1000$. They are overlapping, establishing that our choice of system size, $L$, does not lead to finite size effects in computation of the TDOS and construction of the phase diagram.
\newline    
\par 

\par

\end{document}